\documentclass{PoS}

\PoS{PoS(HEP2005)055}

\def\beq{\begin{equation}}
\def\eeq{\end{equation}}
\def\beqa{\begin{eqnarray}}
\def\eeqa{\end{eqnarray}}
 
\title{Soft-gluon corrections to hard-scattering cross sections through NNNLO}

\ShortTitle{Soft-gluon corrections through NNNLO}

\author{\speaker{Nikolaos Kidonakis}\\
       Kennesaw State University, Physics \#1202, 
       Kennesaw, GA 30144-5591, USA\\
       E-mail: \email{nkidonak@kennesaw.edu}}

\abstract{I present a systematic approach to calculating
soft-gluon corrections through
next-to-next-to-next-to-leading order for arbitrary
hard-scattering cross sections.
Using a unified approach, master formulas are
derived for processes with both simple and complex color flows. 
Applications and numerical results are given
for some processes of interest, including charged
Higgs production and top quark production in the
Standard Model and beyond.}

\FullConference{International Europhysics Conference on High Energy Physics\\
         July 21st - 27th 2005\\
         Lisboa, Portugal}

\begin{document}

\section{Soft-gluon corrections}
Physical cross sections can be calculated in perturbative QCD \cite{QCD} as 
convolutions of  perturbative hard-scattering cross sections ${\hat \sigma}$ 
with non-perturbative parton distribution functions $\phi$.
Near threshold for the production of a specified system 
there is restricted phase space for real gluon emission.
This results in an incomplete cancellation of infrared divergences
between real and virtual graphs, which gives rise to large logarithms.
If we define $s_4=s+t_1+t_2-\sum m^2$, 
where the $t_i$ are standard kinematical invariants and we sum over the 
particle masses, then $s_4 \rightarrow 0$ at threshold  
and the soft and collinear corrections take the form of plus distributions
${\cal D}_l(s_4)\equiv[(\ln^l(s_4/M^2))/s_4]_+$
with $M$ a relevant hard scale and $l \le 2n-1$ for the  
$n$-th order corrections.
The terms with $l=2n-1$ are the leading logarithms (LL), 
those with $l=2n-2$ are the next-to-leading logarithms (NLL),
and so on \cite{KS,KLOS}. These corrections exponentiate in moment space. 
If we define moments of the cross section by
$\hat{\sigma}(N)=\int_0^{\infty} ds_4 \, 
e^{-Ns_4/M^2} \; {\hat\sigma}(s_4)$ then the 
soft corrections become 
${\cal D}_l(s_4)\rightarrow \frac{(-1)^{l+1}}{l+1}\ln^{l+1}N +\cdots$
We can formally resum these logarithms to all orders in $\alpha_s$ by 
factorizing the  soft gluons from the hard scattering.
To invert a resummed cross section back to momentum space one needs to 
employ a prescription to deal with the Landau pole, and this creates 
unavoidable ambiguities. However if one expands the resummed cross section 
at any fixed order, no matter how high, then the inversion can be done 
without resorting to prescriptions \cite{NKuni}. 

At next-to-leading order (NLO) in $\alpha_s$ 
we have ${\cal D}_1(s_4)$ (LL) and ${\cal D}_0(s_4)$ (NLL) terms. 
At next-to-next-to-leading order (NNLO) we have ${\cal D}_3(s_4)$ (LL), 
${\cal D}_2(s_4)$ (NLL), 
${\cal D}_1(s_4)$ (NNLL), and ${\cal D}_0(s_4)$ (NNNLL) terms. 
At next-to-next-to-next-to-leading order (NNNLO) we have 
${\cal D}_5(s_4)$ (LL), ${\cal D}_4(s_4)$ (NLL), 
${\cal D}_3(s_4)$ (NNLL), ${\cal D}_2(s_4)$ (NNNLL), 
${\cal D}_1(s_4)$ (NNNNLL), and ${\cal D}_0(s_4)$ (NNNNNLL) terms. 

Threshold resummation has by now been applied to a large number 
of processes, most recently to 
top quark pair hadroproduction \cite{NKtop},
charged Higgs production \cite{NKchiggs}, 
large-$Q_T$ $W$ \cite{GKS} and Higgs \cite{FKV} production, and 
FCNC top production \cite{NKAB}.
The numerical results typically show that the   
higher-order corrections are sizable and they
dramatically decrease the scale dependence.

\section{NNNLO corrections}
A unified expression for the resummed 
cross section for an arbitrary process is given by \cite{NKuni,NNNLO} 
\beqa
{\hat{\sigma}}^{res}(N) &=&   
\exp\left[ \sum_i E^{f_i}(N_i)+E^{f_i}_{\rm scale}(\mu_F,\mu_R)\right] \; 
\exp\left[ \sum_j {E'}^{f_j}(N_j)\right] 
\nonumber\\ && \times 
{\rm Tr} \left \{H^{f_i f_j} \; 
\exp \left[\int {d\mu' \over \mu'} \Gamma_S^{\dagger\;f_i f_j} \right] \, 
S^{f_i f_j} \; \exp \left[\int  {d\mu' \over \mu'}\; 
\Gamma_S^{f_i f_j} \right] \right\} \, ,
\label{resHS}
\eeqa
where $\mu_F$ is the factorization scale and $\mu_R$ is the 
renormalization scale.
The exponents $E^{f_i}$ and ${E'}^{f_j}$ resum contributions 
from incoming and outgoing partons, respectively, 
$H$ are hard-scattering matrices in color space, and  
$S$ are soft matrices that describe noncollinear soft-gluon emission and whose 
evolution is given in terms of the  soft anomalous dimension matrices 
$\Gamma_S$. See Refs. \cite{NKuni,NNNLO} for details.

The expansion of the resummed cross section to NLO provides us with 
a master formula for the NLO soft-gluon corrections 
\beq
{\hat{\sigma}}^{(1)} = \sigma^B \frac{\alpha_s(\mu_R^2)}{\pi}
\left\{c_3\, {\cal D}_1(s_4) + c_2\,  {\cal D}_0(s_4) 
+c_1\,  \delta(s_4)\right\}
+\frac{\alpha_s^{d_{\alpha_s}+1}(\mu_R^2)}{\pi} 
\left[A^c \, {\cal D}_0(s_4)+T_1^c \, \delta(s_4)\right] 
\eeq
with
$c_3=\sum_i 2 \, C_i -\sum_j C_j$, where 
for quarks $C_q=C_F=(N_c^2-1)/(2N_c)$ and  
for gluons $C_g=C_A=N_c$. 
Also $c_2=c_2^{\mu}+T_2$,
with $c_2^{\mu}=-\sum_i C_i \ln(\mu_F^2/M^2)$ and 
$T_2=- \sum_i [C_i+2\,  C_i \, \ln(-t_i/M^2)
+C_i \, \ln(M^2/s)] -\sum_j [B_j^{(1)}+C_j+C_j \, \ln(M^2/s)]$,
$A^c={\rm tr} \left(H^{(0)} {\Gamma'}_S^{(1)\,\dagger} S^{(0)}
+H^{(0)} S^{(0)} {\Gamma'}_S^{(1)}\right)$, and 
$c_1 =c_1^{\mu} +T_1$, with
$c_1^{\mu}=\sum_i [C_i\, \ln(-t_i/M^2)-\gamma_i^{(1)}]
\ln(\mu_F^2/M^2)+d_{\alpha_s} (\beta_0/4) \ln(\mu_R^2/M^2)$, 
where $B_q^{(1)}=\gamma_q^{(1)}=3 C_F/4$ and $B_g^{(1)}=\gamma_g^{(1)}
=\beta_0/4$.

The master formula for the NNLO soft-gluon corrections is
\beqa
{\hat{\sigma}}^{(2)}&=&
\sigma^B \frac{\alpha_s^2(\mu_R^2)}{\pi^2} \;
\frac{1}{2} \, c_3^2 \; {\cal D}_3(s_4)
\nonumber \\ && \hspace{-25mm}
{}+\sigma^B \frac{\alpha_s^2(\mu_R^2)}{\pi^2} \;
\left\{\frac{3}{2} \, c_3 \, c_2 - \frac{\beta_0}{4} \, c_3
+\sum_j C_j \, \frac{\beta_0}{8}\right\} \; {\cal D}_2(s_4)
+\frac{\alpha_s^{d_{\alpha_s}+2}(\mu_R^2)}{\pi^2} \;
\frac{3}{2} \, c_3 \, A^c\; {\cal D}_2(s_4)
\nonumber \\ && \hspace{-25mm}
{}+\sigma^B \frac{\alpha_s^2(\mu_R^2)}{\pi^2} \; C_{D_1}^{(2)} \;
{\cal D}_1(s_4)
+\frac{\alpha_s^{d_{\alpha_s}+2}(\mu_R^2)}{\pi^2} \;
\left\{\left(2\, c_2-\frac{\beta_0}{2}\right)\, A^c+c_3 \, T_1^c
+F^c\right\} \; {\cal D}_1(s_4)
\nonumber \\ && \hspace{-25mm}
{}+\sigma^B \frac{\alpha_s^2(\mu_R^2)}{\pi^2} \; C_{D_0}^{(2)} \;
{\cal D}_0(s_4)
+\frac{\alpha_s^{d_{\alpha_s}+2}(\mu_R^2)}{\pi^2} \;
\left\{\left[c_1-\zeta_2 \, c_3+\frac{\beta_0}{4} \,
\ln\left(\frac{\mu_R^2}{M^2}\right)
+\frac{\beta_0}{4} \,
\ln\left(\frac{M^2}{s}\right)\right]\, A^c \right.
\nonumber \\ && \hspace{40mm} \left.
{}+\left(c_2-\frac{\beta_0}{2}\right) \, T_1^c 
+F^c \, \ln\left(\frac{M^2}{s}\right)+G^c\right\} \;
{\cal D}_0(s_4) \, .
\eeqa
Here
$C_{D_1}^{(2)}=c_3 \, c_1 +c_2^2
-\zeta_2 \, c_3^2 -(\beta_0/2) \, T_2
+(\beta_0/4) \, c_3 \, \ln(\mu_R^2/M^2)
+c_3\, K/2
-\sum_j (\beta_0/4) \, B_j^{(1)}$,
$F^c={\rm tr} [H^{(0)} (\Gamma_S^{(1)\,\dagger})^2 S^{(0)}
+H^{(0)} S^{(0)} (\Gamma_S^{(1)})^2
+2 H^{(0)} \Gamma_S^{(1)\,\dagger} S^{(0)} \Gamma_S^{(1)}]$, 
and $C_{D_0}^{(2)}$, $G^c$ can be found in \cite{NNNLO} and involve 
two-loop \cite{NK2loop} corrections.

The master formula for the NNNLO soft-gluon corrections is
\beqa
{\hat{\sigma}}^{(3)}&=& 
\sigma^B \frac{\alpha_s^3(\mu_R^2)}{\pi^3} \;  
\frac{1}{8} \, c_3^3 \; {\cal D}_5(s_4)
\nonumber \\ && \hspace{-16mm}
{}+\sigma^B \frac{\alpha_s^3(\mu_R^2)}{\pi^3} \; 
\left\{\frac{5}{8} \, c_3^2 \, c_2 -\frac{5}{2} \, c_3 \, X_3\right\} \;  
{\cal D}_4(s_4)
+\frac{\alpha_s^{d_{\alpha_s}+3}(\mu_R^2)}{\pi^3} \;
\frac{5}{8} \, c_3^2 \, A^c \; {\cal D}_4(s_4)
\nonumber \\ && \hspace{-16mm}
{}+\sigma^B \frac{\alpha_s^3(\mu_R^2)}{\pi^3}  
\left\{c_3 \, c_2^2 +\frac{1}{2}\, c_3^2 \, c_1
-\zeta_2 \, c_3^3 +(\beta_0-4c_2) \, X_3 +2 c_3 \, X_2
-\sum_j C_j \, \frac{\beta_0^2}{48}\right\} {\cal D}_3(s_4)
\nonumber \\ && \hspace{-16mm}
{}+\frac{\alpha_s^{d_{\alpha_s}+3}(\mu_R^2)}{\pi^3} \;
\left\{\frac{1}{2}\, c_3^2\, T_1^c+\left[2\, c_3 \, c_2
-\frac{\beta_0}{2} \, c_3 -4\,  X_3 \right] \,A^c +c_3 \, F^c\right\} \;
{\cal D}_3(s_4)
\nonumber \\ && \hspace{-16mm} 
{}+\sigma^B \frac{\alpha_s^3(\mu_R^2)}{\pi^3} \;
\left\{\frac{3}{2}\, c_3 \,c_2 \, c_1 +\frac{1}{2} \, c_2^3
-3\, \zeta_2 \, c_3^2 \,c_2 +\frac{5}{2} \, \zeta_3 \, c_3^3
+\left(-3 \, c_1+\frac{27}{2} \, \zeta_2 \, c_3\right) \, X_3 
+(3\, c_2-\beta_0) \, X_2 \right.
\nonumber \\ && \hspace{-8mm} \left.
{} -\frac{3}{2} \, c_3 \, X_1-\sum_i C_i \, \frac{\beta_1}{8}
+\sum_j C_j \, \frac{\beta_0}{16} \, 
\left[\beta_0 \, \ln\left(\frac{\mu_R^2}{M^2}\right)+2\, K\right]
+\sum_j \frac{\beta_0^2}{16} \, {B'}_j^{(1)}
+\sum_j\frac{3}{32}\, C_j \, \beta_1 \right\} \; {\cal D}_2(s_4)  
\nonumber \\ && \hspace{-16mm}
{}+\frac{\alpha_s^{d_{\alpha_s}+3}(\mu_R^2)}{\pi^3} \;
\left\{\left(\frac{3}{2}\, c_3 \, c_2-3\, X_3\right) \, T_1^c 
+\frac{3}{2}\, \left[c_2+c_3 \, \ln\left(\frac{M^2}{s}\right)\right]\, F^c
+\frac{3}{2} \, c_3 \, G^c+\frac{1}{2} \, K_3^c \right.
\nonumber \\ && \hspace{-6mm} 
{}+\left[\frac{3}{2}\, c_2^2+\frac{3}{2}\, c_3 \, c_1
-3 \, \zeta_2 \, c_3^2 +3 \, X_2
+\frac{\beta_0^2}{4}-\frac{3}{4}\, \beta_0 \, \left(c_2-\frac{c_3}{2} \, 
\ln\left(\frac{\mu_R^2}{M^2}\right)\right) \right.
\nonumber \\ && \hspace{55mm} \left. \left.
{}-\frac{3\beta_0}{8}\,  c_3\, \ln\left(\frac{M^2}{s}\right)\right] \, A^c
\right\} \;
{\cal D}_2(s_4) + \cdots
\eeqa
where for brevity we have omitted the ${\cal D}_1(s_4)$ and 
${\cal D}_0(s_4)$ terms (full expressions are provided in \cite{NNNLO}).
Here 
$X_3=(\beta_0/12) c_3-\sum_j C_j \beta_0/24$, 
$X_2=-(\beta_0/4)T_2+(\beta_0/8)c_3 \ln(\mu_R^2/M^2)
+c_3K/4-\sum_j(\beta_0/8) B_j^{(1)}$, and 
$X_1, K_3^c$ can be found in \cite{NNNLO}.
 
The NNNLO soft-gluon corrections for top quark production
at the Tevatron in the $q{\bar q}$ channel were calculated in \cite{NNNLO}.
The corrections are small but they decrease further the scale 
dependence of the cross section, which is already small when the 
NNLO soft corrections (for which full results for both partonic channels
and in two different kinematics are available \cite{NKtop}) are taken 
into account. 
When the NNNLO soft corrections are included,  the cross section varies by 
only a few percent as we vary the ratio $\mu/m_t$ by two orders of magnitude.  
The theoretical prediction is in good agreement with data from the Tevatron 
\cite{CDF,D0}.

For charged Higgs production via
$bg \rightarrow tH^-$ the NNNLO soft corrections 
were found to be quite large \cite{NKchiggs,NNNLO}; 
the soft corrections through NNNLO provide an increase of over 80\% over the
leading order cross section at the LHC for a charged Higgs mass 
of 1000 GeV.

\end{document}